# Boosting Perovskite Light-Emitting Diodes Performance by Introducing High Work-function Metal Transition Layer

Laixiang Qin*,

*School of Integrated Circuits, Shanghai Jiao Tong University, Shanghai 200240, China*

Address correspondence to Laixiang Qin, qinlaix@pku.edu.cn

## Abstract

Perovskite light emitting diodes (PeLED) have ignited immense attentions attributed to their marvelous merits of ultra-high color purity, simple fabrication process, low cost and promptly surging device performance. The optimized PeLED device structure conduces to high device performance and longer lifetime which are two main figure of merits that have the last words for their feasibilities of commercialization. Herein, we investigated the impact of the anode's work-function on the performance of an all inorganic PeLED device which was hailed for stability and long lifetime with drift-diffusion and Poisson equations embedded as a computation package in Setfos, a commercially professional software to emulate LED. According to the simulation results, a conclusion that a thin high work-function metal layer inserted between indium tin oxide (ITO) and hole transport layer of nickel oxide ($NiO_x$) substantially facilitated hole injection by decrementing the hole injection barrier. 2 nm thick gold (Au) had been checked and dramatically enhanced luminance intensity by more than 18 fold times had been attained. The situation went for other high work function metals like Pd and Pt as well. Our high work-function thin metal layers decorating anodes in PeLED in facilitating hole injection unfolds a new trajectory towards efficiently boosting PeLED luminance with a technique that is compatible with large scale fabrication process.

## Introduction

Perovskite light emitting diodes (PeLEDs) are recently emerged potential displays devices attributed to their high color purity, a wide color gamut, compatibility with solution process and low cost[1-3]. Ever since perovskite materials had been incorporated into light emitting devices and a low external quantum efficiency (EQE) of 0.1 % had been obtained in green PeLED with a device structure of $ITO/PEDOT:PSS/CH_3NH_3PbBr_3/F8/Ca/Ag$ and a EQE of 0.76 % had been obtained in infrared radiance with perovskite emitter $CH_3NH_3PbBr_2I$ superseding $CH_3NH_3PbBr_3$ in the PeLED device structure mentioned above in 2014[4], PeLEDs have witnessed EQE proliferating over the past years, with the green, red and blue PeLEDs having obtained over 30%, 25% EQE, demonstrating monumental potentials to be used in display or lightning devices[5-8]. Thanks to the ultra-narrow full width at half maximum (FWHM) ( $\leqslant$ 20 nm) generally obtained in PeLED compared with organic light emitting diode (OLED) (FWHM generally larger than 45 nm) and quantum dots based LED (FWHM is around 30 nm), PeLED demonstrates enormous application potentials in high resolution and high angle near-eye displays[9-12]. Whereas, the commercialization road of PeLEDs has been encountered with major challenges of shorter lifetime that is far from fulfilling the requirements of commercial luminescent products and smaller EQE. The commercialization of PeLEDs largely hinges on their high stability as well as superior EQE, to which substantial attentions have been garnered among researchers to devote endeavors to[13-15]. It is notable that the crystalline quality of perovskite materials largely determine the performance of PeLED, moreover, the device structure of PeLED have the last words for the device performance of PeLED, thus it is of importance to delicately devise the structure of PeLED to obtain high device performance[16-20]. Herein, we devised an all inorganic PeLED with a device structure of $ITO/NiO_x$/Perovskite/ZnO/LiF/Al to obtain high stability by making use of the environmental stability characterization of inorganic hole transport (HTL) and electron transport layer (ETL). The devised PeLED was investigated by simulation method by solving the drift-diffusion, Poisson and current continuity equations which had been embedded in the commercial Setfos software as a computation package. The electric field, potential and charge carrier distributions were demonstrated to have profound effect on the performance of PeLED. According to the simulation results, it was found that the electrode injection barrier fundamentally determined the current density in the light emitting device, and in this device structure, a thin high work-function metal layer inserted between ITO anode and HTL showcased a hole injection boosting phenomenon which was manifested by abrupt soaring in current density and luminance intensity. Our work sheds light on substantially increasing PeLED luminance intensity with an feasible gateway of inserting a thin high work-function metal layer, demonstrating the instructive merits of simulation method in devising high performance PeLED device.

## Results and Discussion

The simulated all inorganic perovskite LEDs with and without thin high work-function metal layer are depicted in Fig.1 (a) and (b). The relative energy diagrams are demonstrated in Fig.1 (c) and (d). The control PeLED device possesses the structure of ITO (100 nm) /$NiO_x$ (20 nm)/Perovskite (100 nm)/ZnO (20 nm)/LiF (2 nm)/Al (100 nm) (referred to as control-PeLED) while the thin layer Au decorated PeLED is featured as ITO (100 nm) /Au (2 nm)/$NiO_x$ (20 nm)/Perovskite (100 nm)/ZnO (20 nm)/LiF (2 nm)/Al (100 nm), which is termed as Au-PeLED. The simulation block adopted is depicted from equation (1) to (5), where drift-diffusion,

Poisson and continuity equations embedded in Setfos as a computation package are used to describe the carrier transport, carrier distributions, energy barrier and electric field distributions of both control- and Au-PeLEDs to corroborate the hole injection boosting up behavior of inserted Au on ITO anode. For simplicity, the interfaces are treated as ideal case and the ions migration is neglected as well. In the presence of a constant E, drift-diffusion equation can be depicted as,

$$\begin{aligned} J_n &= J_{n-drift} + J_{n-diff} = qn\mu_n E + q\frac{dn}{dx}D_n \\ J_p &= J_{p-drift} + J_{p-diff} = qp\mu_p E - q\frac{dp}{dx}D_p \end{aligned} \tag{1}$$

Where $J_n$, $J_p$, $J_{n\text{-drift}}$, $J_{p\text{-drift}}$, and $J_{n\text{-diff}}$, $J_{p\text{-diff}}$ represent total electron/hole current densities, electron/hole drift current densities, and electrons/holes diffusion current densities respectively. n and p are the electron and hole densities. $\mu_n$, $\mu_p$ denote the electron and hole mobility. q depicts the elementary charge. $D_n$ and $D_p$ are the diffusion coefficients of electrons and holes. When the drift and diffusion currents are in balance and recombination or generation are absent, the total $J_n$ and $J_p$ equal 0, and supposing electrons and holes are confined in the energy barrier x=0 and x=a respectively, according to Einstein's relation that $D = \mu\frac{kT}{q}$ , one can obtain the solution that

$$\begin{aligned} n &= n_0 e^{\frac{-x}{L}} \\ p &= p_0 e^{-\frac{a-x}{L}} \end{aligned} \tag{2}$$

Where $n_0$ and $p_0$ are the initial electrons and holes density, and L can be described as,

$$L = \frac{kT}{qE} \tag{3}$$

While on the other hand, when recombination R is in present, the continuity equation for electrons and holes and Poisson equation that describes the carrier currents and distribution are demonstrated as below,

$$\begin{aligned} q\frac{\partial n}{\partial t} &= \nabla \cdot J_n - R \cdot q \\ q\frac{\partial p}{\partial t} &= -\nabla \cdot J_p - R \cdot q \end{aligned} \tag{4}$$

$$\nabla \cdot (-\varepsilon_r \nabla V) = q(p - n + N_d^+ - N_a^-) \tag{5}$$

where $\varepsilon_r$ is referred to as the relative permittivity. $N_d^+$, $N_a^-$ represent the donor and acceptor densities respectively.

**I 2nm Au on ITO facilitating hole injection**

The thin Au metal layer was introduced between ITO and HTL $NiO_x$ to facilitate hole injection from anode to HTL attributed to higher injection barrier (0.5 eV) between ITO anode and $NiO_x$ valence band maximum (VBM). And moreover, 2 nm Au is so thin that it will does little effect on the absorption, reflection and transmission behaviors of the PeLED device structures, as demonstrated in Fig.2 (a), (b) and (c), where the absorption, reflection and transmission curves of Au-PeLEDs showcase little variations from those of control-PeLEDs. Fig.2(d) showcased the transmittance curve of a passive structure of Au/(100nm) ITO, where an evident reduction in transmittance could be observed as Au grew over 10 nm thickness. Whereas, for another, the work function of Au is 5.1 eV, much higher than 4.7 eV of ITO, as depicted in Fig.1(d). The incorporation of 2 nm Au on ITO successfully augments the work function of the combined anode from 4.7 eV to 5.1 eV, which aligns well with the VBM of $NiO_x$ of 5.2 eV, leading to an ohmic contact and sufficient hole injection between the combined anode and $NiO_x$ HTL, resulting in a boosting of luminance efficiency in Au-PeLED with regarding to that of the control-PeLED by more than 18-folds under a bias voltage of 3 V, as shown in Fig.3(a). When sweeping the biased voltage from 0.5 to 3 V, the luminance intensities and efficiencies of Au-PeLED all lies well above those of the control device, indicating the efficacy of the inserted 2 nm Au metal layer in boosting the PeLED performance via facilitating hole injection.

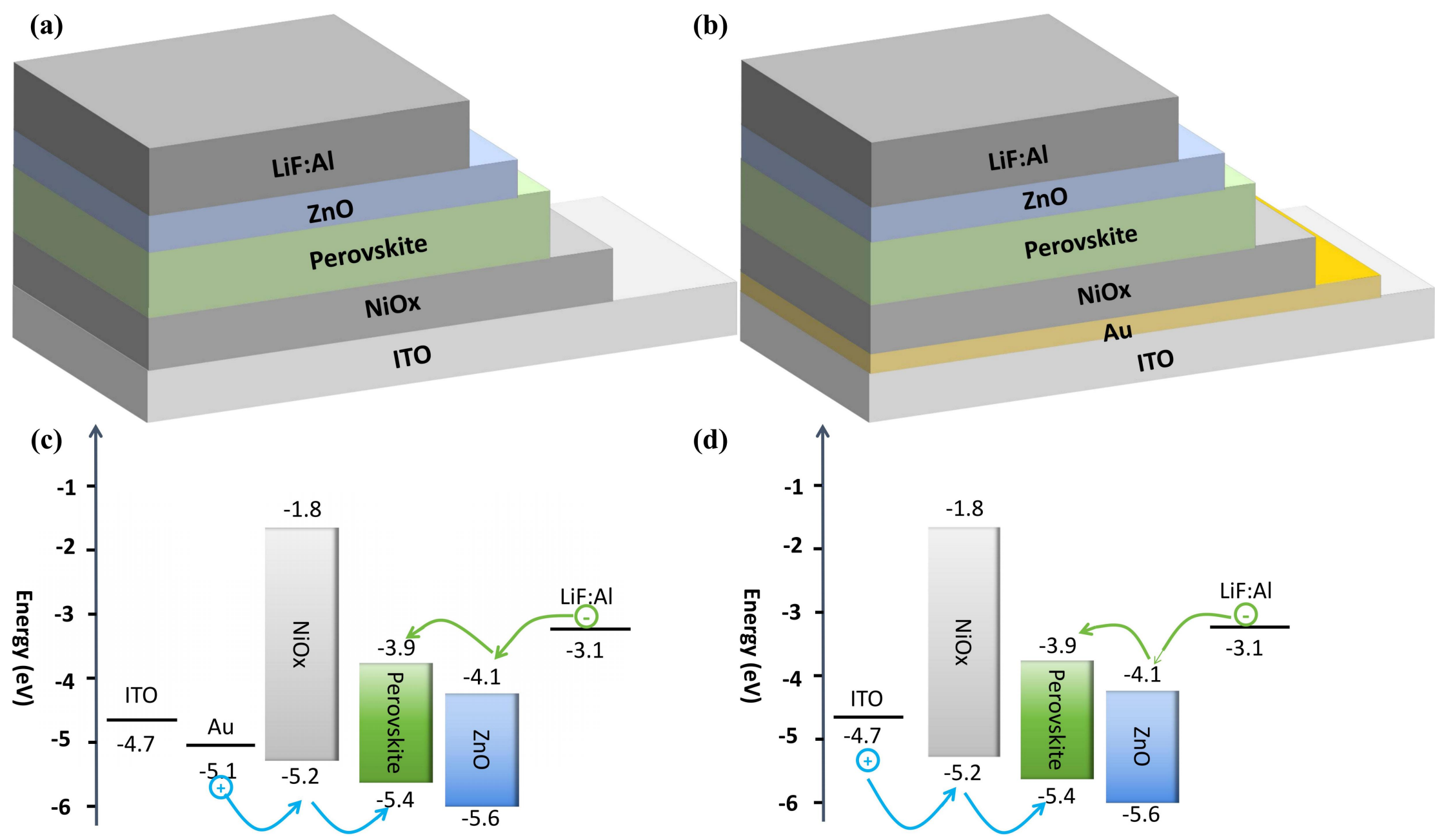

**Fig.1** The simulated device structures of all inorganic PeLEDs with (a) and without (b) thin Au metal layer on ITO; Band diagrams of simulated PeLEDs with (c) and without (d) thin Au layer on ITO.

The impacts of Au thickness variations deposited on ITO on the luminance of Au-PeLEDs had been checked from 0.5 to 20 nm, as demonstrated in Fig.3 (c) and (d), from which the conclusion can be drawn that the luminance intensity attenuates with Au thickness increment attributed to the enhanced absorbance and reflectance of thick Au, as shown in Fig.2 (d), (e) and (f) that show the absorption, reflection and transmittance maps of Au/ITO/glass with different Au thickness. Whereas, the light extraction loss caused by thick Au can be compromised by the emission boosting caused by surged holes injection, resulting in enhanced light emission as well, which demonstrating the robustness of the proposed avenue to improve the performance of all inorganic PeLED. As a matter of fact, the emission intensity of Au PeLED can still maintain more than half of its original emission intensity when Au grows from 2 nm to 20 nm though the transmittance of 20 nm Au decorated PeLED has greatly degraded, manifested by the less than 10% transmittance of the passive structure of Au (20 nm)/ITO (100 nm), at its emission peaks of 520 nm, as shown in Fig.2(f). Whereas, intriguingly, the thickness of Au layer does no affect on recombination and electric field distribution, as shown in Fig.3 (e) and (f), where the recombination and electric field distribution curves are totally coincident with each other as Au layer varies from 0.5 to 20 nm. That is to say, the hole injection improvement obtained in Au-PeLED is totally derived from high work function modification of anode caused by metal Au, as long as Au can form a continuous layer, the thinner the Au layer, the better the efficacy is attributed to less absorption and reflectance leading to less light outcoupling loss.

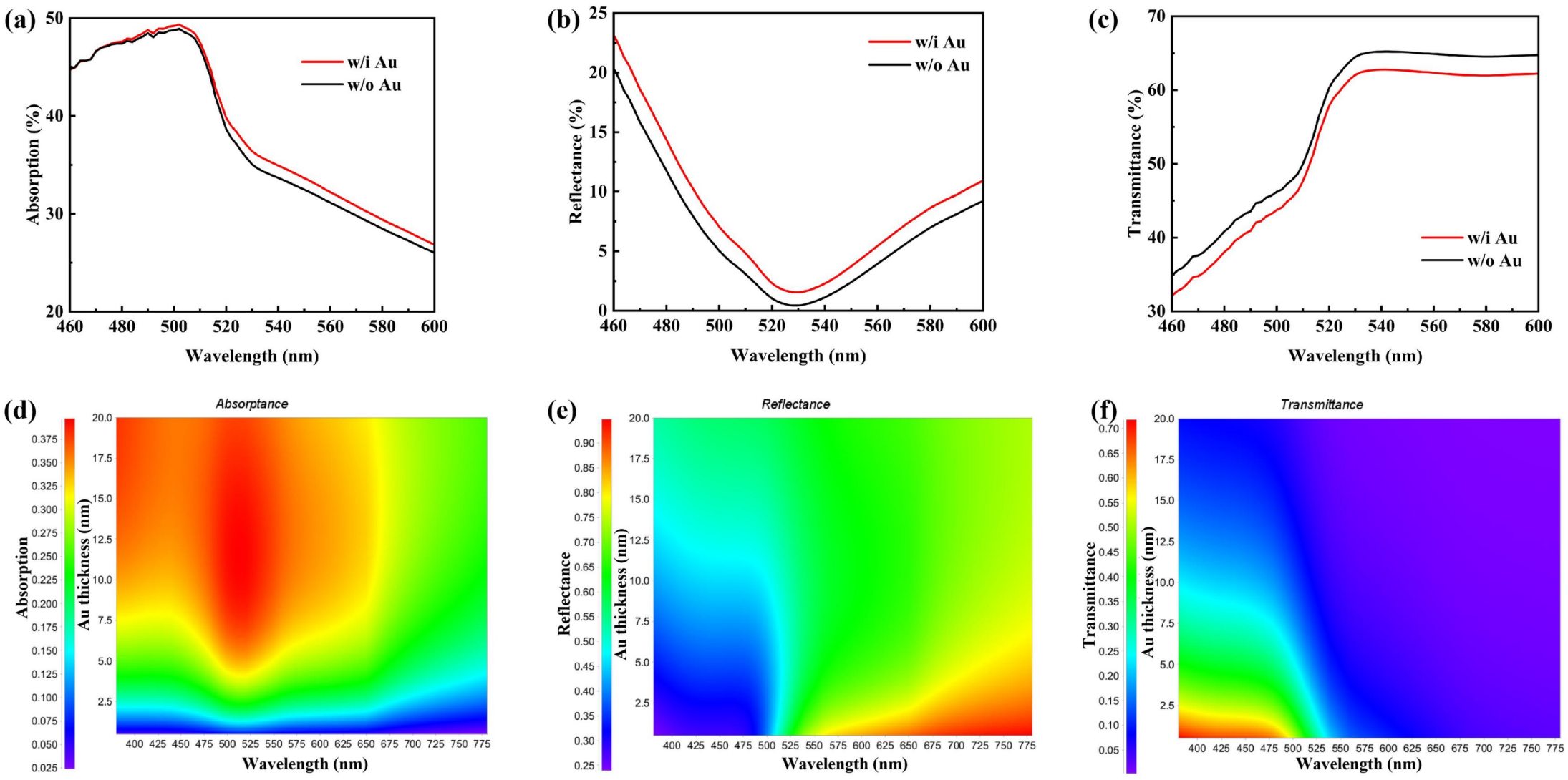

**Fig.2** The absorption (a), reflection (b) and transmittance (c) curves of Au-PeLED (the red lines) in compared with those of control-PeLED (the dark lines), w/i and w/o refer to with and without respectively. The absorption (d), reflectance (d) and transmittance (e) maps of Au/ (100 nm) ITO/Glass structure with Au thickness varying from 0.5 to 20 nm.

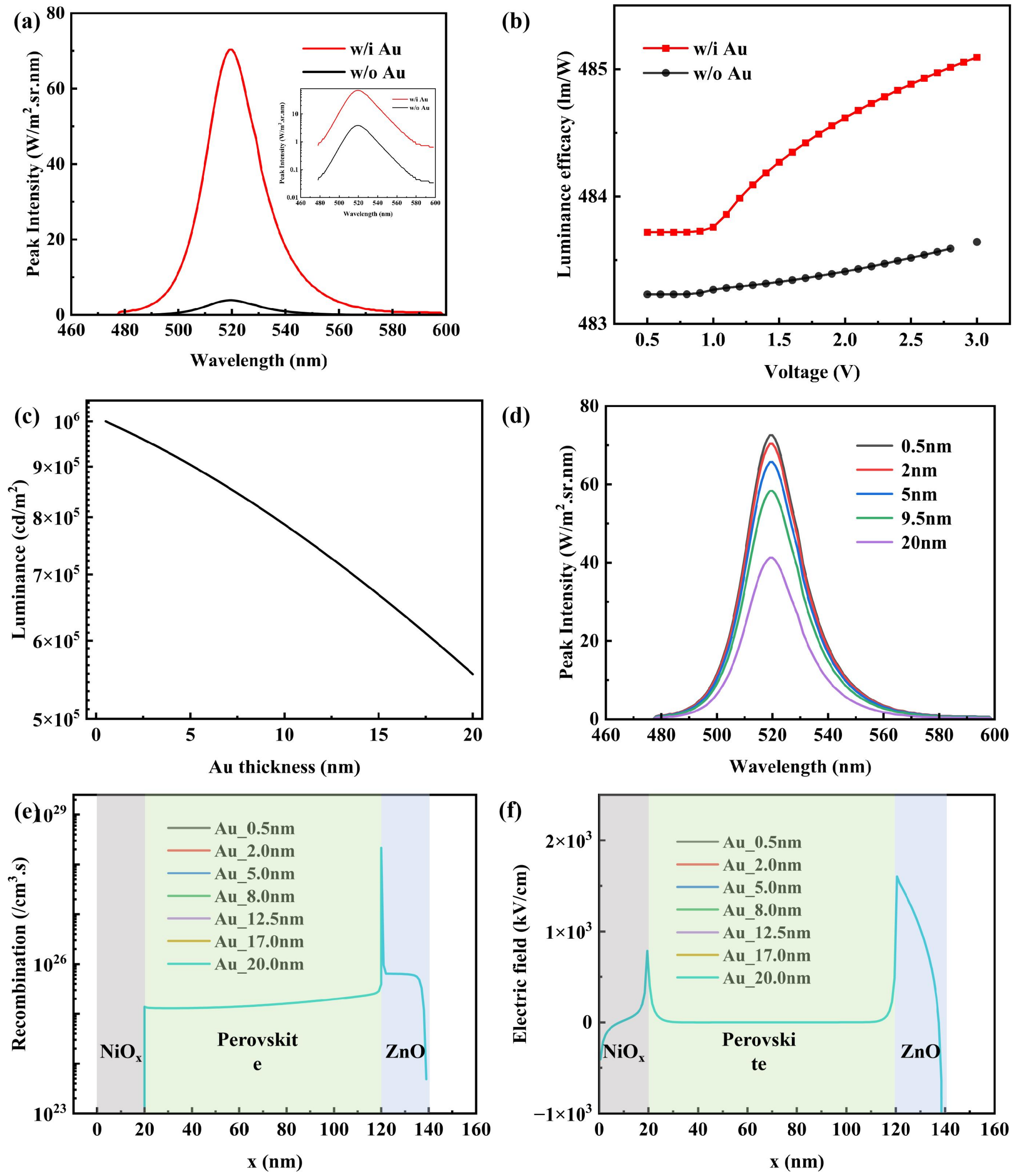

**Fig.3** (a) The emission curves comparison between Au- and control- devices under 3 V voltage bias; luminance efficiency vs applied voltages in Au-PeLED (the red lines) in compared with those of control-PeLED (the dark lines). The inserted figure demonstrates the emission curves in the logarithmic coordinates; (c) the emission peak intensity versus Au thickness for Au-PeLED under voltage bias of 3 V; (d) luminance curves of Au-PeLEDs with regarding to different inserted Au layers; Recombination (e) and electric field (f) curves of Au-PeLED with different Au thickness, demonstrating totally coincident curves when Au layer varies from 0.5 to 20 nm, signifying the hole injection boosting phenomenon totally arising from the work function of metal Au.

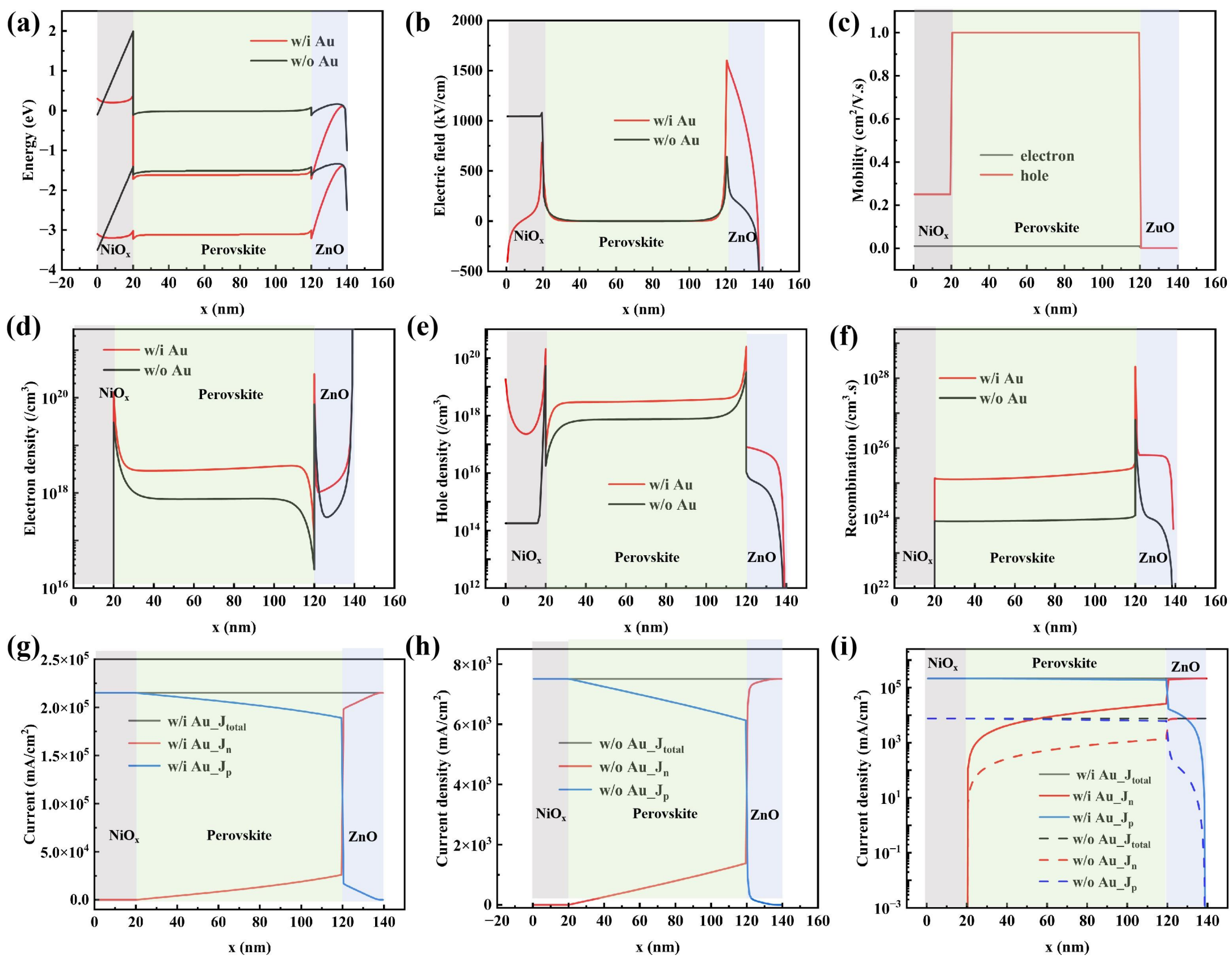

**Fig.4** Energy band diagrams (a), electric fields (b), mobility (c), electron density (d), hole density (e) and recombination distribution curves (f) under voltage bias of 3 V for Au- (the red lines) and control-PeLEDs (the dark lines); current density distribution curves of (g) control- and (h) Au-PeLEDs and (i) current density distributions comparison between control- and Au-PeLED under voltage bias of 3 V.

To further elucidate the above phenomena, the energy band diagrams, carrier mobility, electric field, carrier density and recombination distribution curves as well as current density curves of control- and Au-PeLEDs are depicted in Fig.4. From Fig.4(a), it can be concluded that the injection energy barrier for holes in Au-PeLED has been immensely reduced by 0.4 eV from 0.5 eV to 0.1 eV with the introduction of a thin Au layer, whereas, the situation is different in control-PeLED where the holes are encountered with higher injection barriers (0.5 eV), resulting in higher hole density in Au-PeLED in compared with control-PeLED as shown in Fig.4 (e), which can be explained by the carrier injection equation as showcased below,

$$p_{inj} = p_{anode} e^{-\frac{\Delta E_V}{kT}} \tag{6}$$

The injected holes $p_{inj}$ density exponentially contingents upon the hole injection energy barrier $\Delta E_V$, thus a reduction of $\Delta E_V$ will monumentally boost the hole injection density, that is why the insertion of a thin layer Au has escalated the hole injection substantially by reducing the hole injection barrier from 0.5 eV to 0.1 eV, which significantly increases the hole injection, as demonstrated in Fig.4 (e) with Au-PeLED possessing immensely enhanced hole current density under 3 V voltage bias. For another, the band diagram for electrons at 3 V bias in Au-PeLED favors the electrons injections attributed to hole accumulation leading to band energy shifting down, facilitating electrons injection as well, resulting in higher electron density in Au-PeLED in compared with control-PeLED, as demonstrated in Fig.4 (d). The two merits brought about by thin layer Au insertion cumulatively lead to higher recombination and luminance in Au-PeLED, as showcased in Fig.4 (f). The favored electron injection at higher voltage bias can also be withdrawn from Fig.4 (b), where a larger electric field peak exists in Au-PeLED in compared with control PeLED, which facilitates electron injection, consistent with the phenomenon observed in Fig.4 (d). There are electron distribution peaks in the heterojunction region between ETL and emission layer in both Au- and control-PeLEDs attributed to the 0.2 eV injection barrier between ZnO and perovskite emission layer, where the electron distribution peak in the Au-PeLED is higher than that of the control-PeLED thanks to the favorable energy bandgap for electron injection at 3 V bias. The incorporation of thin Au layer into anode leads to a sharp increment in hole distribution density curve in compared with PeLED without Au layer. The hole density distribution peaks lying between the emission layer and ETL derive from the higher hole mobility than electron mobility. From the recombination distribution curves in Fig.4 (f), there are emission peaks between emission layer and the ETL attributed to large electrons and holes accumulation and emissive recombination. Attributed to the higher hole mobility in HTL in compared with electron mobility in ETL, emission mainly occurs in perovskite in proximity to ETL, as demonstrated in Fig.4 (c), which can be alleviated by doping ZnO to improve its electron mobility to align with the hole mobility in $NiO_x$ or to reduce the ETL layer thickness. Fig.4 (g), (h) and (i) summarize the current density distribution curves of control- and Au- PeLEDs, from which one can see that the introduction of 2 nm Au between HTL and ITO anode can significantly escalate the current density in Au-PeLED by more than one order of magnitude compared with that of control-PeLED, further substantiating the hole injection boosting effect brought about by the thin Au layer on ITO.

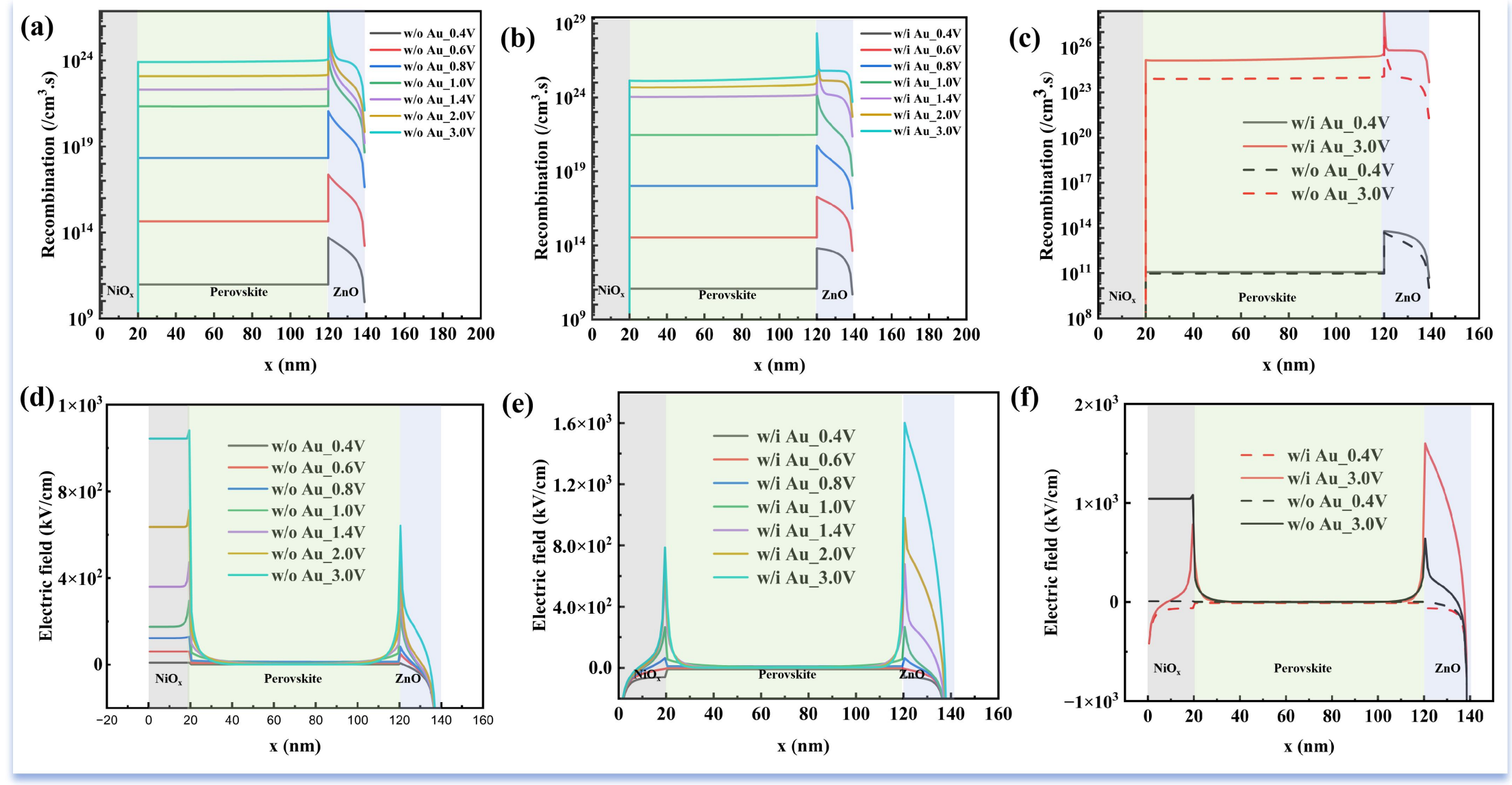

**Fig.5** The recombination distribution curves of control- (a) and Au-PeLEDs (b) under different voltage biases; (c) the recombination distribution curves comparisons between control- and Au-PeLEDs under voltage bias of 0.4 and 3 V respectively; electric field distribution curves of control- (d) and Au-PeLEDs (e) under different voltage biases; (f) the electric field distribution curves comparisons between control- and Au-PeLEDs under voltage bias of 0.4 and 3 V respectively.

Fig.5 depicts the recombination and electric field distribution curves under different applied biasing voltages. In both control- and Au-PeLEDs, recombination increases monumentally with the increase of voltage, indicating enhanced electron/hole injections with higher applied voltage, resulting in higher recombination intensities, as demonstrated in Fig.5(a), (b). According to the recombination equation given below,

$$R = B \cdot n \cdot p \propto n_{inj} \cdot p_{inj} \tag{7}$$

R depends on the electron and hole injection density, thus the significantly enhanced holes and electrons injections in Au-PeLED account for the recombination surging compared with that of the control-PeLED as demonstrated in Fig.5(c), from which it can be concluded that when 2 nm Au thin metal layer has been inserted between the ITO and the $NiO_x$ layer, the recombination increases by more than one-order of magnitude attributed to the modification effect of Au on the work function of anode, leading to a boosting of hole injection, which contributes to more holes and electrons to recombine with each other, leading to more radiative recombinations in Au-PeLED at all voltages applied. The recombination peaks in the control- and Au-PeLEDs under all supplied voltages lying in proximity to ETLs arise from the imbalanced electrons and holes mobility as shown in Fig.4 (c), which leading to charges recombining mainly between the interfaces of emission perovskite layer and ZnO ETL. Moreover, the holes injection boosting caused by the incorporation of 2 nm thick Au on ITO can also be consolidated by the electric fields distribution curves as showcased in Fig.5 (d), (e). From the electric field curves of PeLED with Au, the electric field between anode and HTL demonstrates a lower height compared with those of control device, indicating the inserted Au layer contributing to favorable hole injection. For another, the electric field between ETL and cathode in thin Au layer modified PeLED demonstrated higher peaks than those of control device, which indicates that the electron injection between cathode and ETL has also been enhanced attributed to more holes can be injected in the Au-PeLED, finally resulting in improved recombination and emission in PeLED with Au thin metal layer. Thus, as a whole, the introduction of a thin metal layer between anode and HTL can not only boost hole injection, but also facilitate electron injections, all of which contribute to high recombination in perovskite emission layer, resulting in enormously surged luminance as showcased in Fig.3.

As for the luminance intensity of Au-PeLED in accordance with Au thickness demonstrated in Figure 3 (c) and (d), the intensity droop according to the increasing of Au thickness is mainly attributed to the degraded transmittance of thicker Au, as demonstrated in Figure 3 (f). Attributed to the deteriorated absorption and increased reflectance, part of the emitted light has been absorbed, part has been reflected into the inner part of the PeLED device, and what is more, surface plasma polarization sacrifice or metal electrode induced quenching may happen as well, resulting in significant outcoupling light loss in PeLED with thicker Au interlayer. The electric field distribution and recombination curves with different Au inserting layer demonstrate no difference, as shown in Fig.3 (e) and (f), a phenomenon that can sufficiently substantiate the effectiveness of Au layer in boosting the hole injection of Au-PeLED, which has no dependence on the thickness of the Au layer, only the work function matters.

## Ⅱ The effects of various high work-function metals

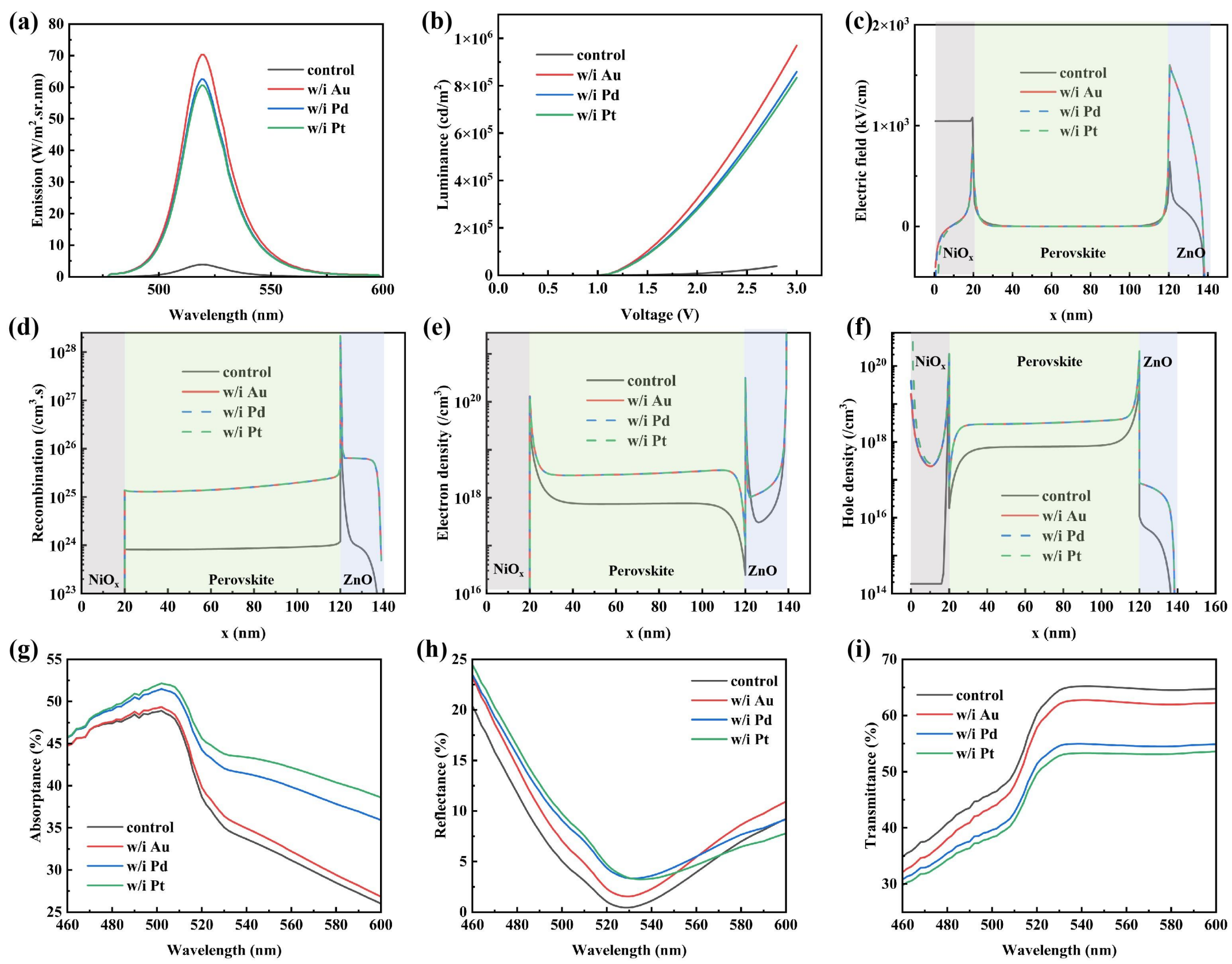

**Fig.6** The peak luminance intensity under 3V voltage bias (a), luminance under different voltages (b), electric field (c), recombination (d), electron densities (e) and hole densities distribution curves of control- and different high work function metals decorated PeLEDs under bias voltage of 3 V; the absorption (g), reflectance (h) and transmittance (i) curves of control- and different high work function metals decorated PeLEDs.

It is natural to be brought to the conjecture that other kinds of high work function metals similar to Au have the same effect on PeLED with the structure of ITO/$NiO_x$/perovskite/ZnO/LiF/Al. To work out this conundrum, we replace a thin layer Au with high work function metals of Pd and Pt which are characterized of high work function of 5.12 and 5.65 eV respectively. The simulation results are showcased in Fig.6. Impressively, the luminance intensity increases by more than 16- and 15-times than that of the control-PeLED that has no interlayer of high work function metal layer but smaller than the luminance intensity of that of Au modified PeLED, as demonstrated in Fig.6 (a) and (b). Since all three kinds of inserted metal layers have higher work functions that are feasible to effectively inject holes from anode to the HTL, thus the electric field and recombination distribution curves will be more or less the same since they mainly depend on band energy alignments, which is consistent with the phenomenon showcased in Fig.6 (c)-(e), where the recombination and electric field and electron density distribution curves demonstrate marginal difference among three kinds of high work function metals decorated PeLEDs, while all three curves are monumentally larger than that of the control PeLED, indicating the superior hole injection boosting behaviors brought about by the incorporation of high work function metals between anodes and HTLs, as shown in Fig.6 (c), (d) and (e). The hole density distribution curves present a little difference among three kinds of high work function metals decorated PeLEDs, with the hole density of Pt based PeLED is the largest among the three kinds of PeLEDs, and Pd based PeLED demonstrates hole density distribution curves larger than that of Au based PeLED. The phenomenon is due to the higher work functions of Pt (5.65 eV) and Pd (5.12 eV) than that of Au (5.1 eV) facilitate to more holes injection. Similarly, the hole density distribution curves of three kinds of high work function metals decorated PeLEDs also showcase significant surging compared with that of the control-PeLED, solidifying the enhanced hole injection caused by high work function metal decorating ITO anode. Now that the recombination curves among the three kinds of high work function metals decorated PeLEDs show little difference, the emission peak intensity difference among the three kinds of PeLEDs inserted with high work function metals stems from the different optical properties of 2 nm thin metal films, as demonstrated in Fig.6 (g), (h), (i), where the absorption, reflection and transmittance curves of PeLEDs with 2 nm Au, Pd, Pt and without any metals have been presented. From the curves mentioned above, the conclusion can be drawn that the control-PeLED demonstrates the best optical properties for emitted light generated in the perovskite emissive layer to be extracted, followed by that of the Au decorated PeLED, and the Pt decorated PeLED show the infaustest optical properties for emitted light to be extracted originating from the most non-transparent Pt film, as demonstrated in Fig.6 (i), which can be accounted for by the highest absorptive and reflective coefficients, as depicted in Fig.6 (g) and (h).

## Conclusion

In conclusion, we have put up with a novel avenue to enormously boost the hole injection in PeLED via a feasible and large-scale-fabrication compatible high work function thin metal layer depositing technique. By incorporating a 2 nm thin Au layer on ITO anode, the hole injection has been dramatically improved, consolidated by the prominent emission intensities obtained in high work function metals decorated PeLEDs in compared with that of the control device, where more than 18-, 16- and 15- folds emission peak intensities increments have been obtained in 2 nm thick Au, Pd and Pt decorated PeLEDs with regards to that of the control-PeLED. The boosting in emission intensities totally derives from the enhanced carriers injection, compromising of holes and electrons injection simultaneously, where holes injection increment is brought about directly by higher work function metals modified anodes, while the electron injection escalation is an accessory advantage of boosted hole injection. The proliferated electrons and holes injection caused by high work function metal layer insertion cumulatively contribute to significant enhanced recombination, thus resulting in substantially augmented luminance intensity. The luminance intensity grows weaker when Au is thicker due to deteriorated transmittance of thick Au layer which lead to degraded light extraction feasibility of PeLED. On the other hand, the recombination and electric field distribution curves remain immutable as the Au grows thicker, attesting to the perspective that it is the work function that is pivotal to the hole injection increment phenomenon observed in high work function metals decorated PeLEDs and thus the thinner the metal layer the better as long as a continue metal layer can be formed since the inserted metal layer will block and reflect the emitted light when metal layer grows too thick. In all, the proposed boosting hole injection in PeLED via high work function metal decorating ITO anode is feasible and compatible with large scale synthesis, paving the way for escalating the emission intensity drastically with large scale synthesis compatible technique, expediting the commercialization of PeLEDs.

## Data availability

Data that support the findings of this study are available from the corresponding author upon reasonable request. Source data are provided with this paper

## Acknowledgement

This work was supported by the National Natural Science Foundation of China (nos. 62475151 and 62074098 to Changxin Chen). We thank the Center for Advanced Electronic Materials and Devices (AEMD) at Shanghai Jiao Tong University.